\documentclass[10pt,a4paper]{article}
\usepackage{graphicx}
\usepackage{subfigure}
\usepackage{amsmath,amsthm}
\usepackage{amssymb,amsfonts}
\usepackage{authblk}
\usepackage{url}
\usepackage{hyperref}

\newtheorem{definition}{\textbf{Definition}}

\title{Testing Copula Hypothesis with Copula Entropy}
\author{Jian MA\thanks{Email: majian03@gmail.com}}
\affil{Hitachi China Research Laboratory}

\begin{document}

\maketitle

\begin{abstract}
	\noindent
Testing copula hypothesis is of fundamental importance in the applications of copula theory. In this paper we proposed a copula hypothesis testing with copula entropy. Since copula entropy is a unified theory in probability and therefore testing copula hypothesis based on it can be applied to any types of copula function. The test statistic is defined as the difference of copula entropy of copula hypothesis and true copula entropy. We propose the estimation method of the proposed statistic and two special cases for Gaussian copula hypothesis and Gumbel copula hypothesis. We test the effectiveness of the proposed method with simulation experiments.
\end{abstract}

\section{Introduction}
Evaluating the fitness of models to data is a common practice in scientific activities. Testing hypothesis is one of the fundamental problems in statistics and and has wide applications in every branch of sciences. 

Copula theory is about representing multivariate dependence with copula functions \cite{nelsen2007,joe2014}. As the core result of copula theory, Sklar's theorem \cite{sklar1959} states that multivariate density function can be represented as a copula function with marginal functions as its inputs. There are many copula function families available for real applications, such as Gaussian copula, t Copula \cite{Demarta2005}, Archimedean copula, Archimax copula \cite{Mesiar2013}, Sibuya Copula \cite{Hofert2013}, among others. 

Modeling with copula function is a widely used methods in many scientific fields \cite{Durrleman2000,Genest2009,Patton2012,Fan2014,Tootoonchi2022,NazeriTahroudi2023} and hence testing copula hypothesis is important in those practices \cite{Berg2013}. Many research have been contributed to copula hypothesis testing, such as testing Gaussian copula hypothesis \cite{Malevergne2003,Amengual2020}, testing Archimedeanity \cite{Jaworski2010,Buecher2012}, testing symmetry of copula \cite{Genest2012}. Current work on testing copula hypothesis are mainly focusing on special types of copula function and a general method for any types of copula is needed.

Copula Entropy (CE) is a recently proposed theory in probability. It defined the concept of CE as a special kind of Shannon entropy with copula function \cite{Ma2011}. Copula function represents the dependence relationship between random variables while CE measures such relationship in a unified way. Contrast to other dependence measures based on copula, such as Spearman's $\rho$ and Kendall's $\tau$, CE has many good properties, including non-negative, invariance to monotonic transformation, and equivalent to correlation matrix under Gaussianity. CE has been applied to hypothesis testing recently, including multivariate normality test \cite{Ma2022}, two-sample test \cite{Ma2023b}, change point detection \cite{Ma2024a}, and symmetry test \cite{Ma2025a}.

In this paper, we proposed a copula hypothesis testing with copula entropy. It can be used for any types of copula hypothesis testing. The test statistic is defined as the difference of CE of copula hypothesis and true CE. We give the estimation method of the proposed statistic and two cases for Gaussian copula hypothesis and Gumbel copula hypothesis. We test the effectiveness of the proposed method with simulation experiments.

This paper is organized as follows: Section \ref{sec:ce} introduces the basic theory of CE, Section \ref{sec:test} presents the proposed testing method, Section \ref{sec:est} gives the estimation method of the proposed statistic, simulation experiments will be presented in Section \ref{sec:sim}, Section \ref{sec:con} concludes the paper.

\section{Copula Entropy}
\label{sec:ce}
With copula theory, Ma and Sun \cite{Ma2011} defined the concept of Copula Entropy as follows:
\begin{definition}[Copula Entropy]
	Let $\mathbf{X}$ be random variables with marginals $\mathbf{u}$ and copula density function $c$. The CE of $\mathbf{X}$ is defined as
	\begin{equation}
	H_c(\mathbf{x})=-\int_{\mathbf{u}}{c(\mathbf{u})\log c(\mathbf{u})d\mathbf{u}}.
	\label{eq:ce}
	\end{equation}	
\end{definition}

They also proposed a non-parametric estimator of CE \cite{Ma2011} comprising of two simple steps:
\begin{enumerate}
	\item estimating empirical copula density function with rank statistic;
	\item estimating the entropy of the estimated empirical copula density with the kNN entropy estimator\cite{Kraskov2004}.
\end{enumerate}

If the copula density function $c$ is given, the CE can also be estimated with the following way:
\begin{equation}
	H_c(\mathbf{x})=-E(\log c(\mathbf{u})).
\end{equation}

\section{Testing Copula Hypothesis}
\label{sec:test}
Given random variables $\mathbf{X}\in R^n$ and its samples $\mathbf{X}_T$ associated with copula density function $c_\mathbf{x}(\mathbf{u})$. Our goal is to test whether $c$ belong to a hypothesis $c(\mathbf{u})$, the null hypothesis of the problem is
\begin{equation}
	H_0: c_\mathbf{x}(\mathbf{u}) = c(\mathbf{u});
\end{equation}
alternative hypothesis is
\begin{equation}
	H_0: c_\mathbf{x}(\mathbf{u}) \neq c(\mathbf{u}).
\end{equation}

We propose to test copula hypothesis with CE. The principle of testing is to compare the CE of the copula hypothesis with the true CE:
\begin{equation}
	T_c(\mathbf{X}_T|c)=H_c(\mathbf{X}_T|c)-H_c(\mathbf{X}_T|c_\mathbf{x}),
	\label{eq:stats}
\end{equation}
The first term is the CE under the hypothesis of copula $c$ and the second term is true CE. If $H_0$ is true, then $T_c$ should be 0; otherwise, $T_c$ should be large value.

Since CE is a unified theory for copula function, testing copula hypothesis based on CE can be used for any types of copula function. The only work needed is to select the copula family $c$.

\section{Estimation}
\label{sec:est}
The statistic in \eqref{eq:stats} can be estimated as two part. The second term is true CE and therefore can be estimated directly from data with the nonparametric estimator of CE. The first term is the CE of copula hypothesis which can be estimated in the following 3 steps:
\begin{enumerate}
	\item estimate empirical copula density $\mathbf{\hat{u}}$ from $\mathbf{X}_T$;
	\item estimate the parameters $\alpha$ of copula $c$ with $\mathbf{\hat{u}}$;
	\item calculate the CE of the copula hypothesis with the following equation:
	\begin{equation}
		H_c(\mathbf{X}_T|c)=-E(\log c(\mathbf{\hat{u}};\alpha)).
		\label{eq:celikelihood}
	\end{equation}
	
\end{enumerate}

In the first step, empirical copula density can be estimated with rank statistic and in the second step, the parameters $\alpha$ of copula density can be estimated with the likelihood method \cite{joe2014}.

Here we give two special cases of estimating CE of copula hypothesis:

\paragraph{Gaussian Copula}
Gaussian copula density function can be written as the following \cite{Arbenz2013}:
\begin{equation}
	c_n(\mathbf{u})={|\Sigma_\rho|^{-\frac{1}{2}}}exp\left\{-\frac{1}{2}\Phi(\mathbf{u})(\Sigma_\rho^{-1}-I)\Phi^T(\mathbf{u})\right\},
	\label{eq:igc}
\end{equation}
where $\Sigma_\rho$ is correlation matrix, $\Phi$ is quantile normal function, and $I$ is identity matrix.

So estimating CE of Gaussian copula hypothesis can be done as follows:
\begin{enumerate}
	\item estimate $\Sigma_\rho$ from $\mathbf{X}_T$;
	\item calculate the value of Gaussian copula with \ref{eq:igc};
	\item calculate CE of copula hypothesis with \ref{eq:celikelihood}.
\end{enumerate}

\paragraph{Gumbel Copula}
Gumbel Copula is one member of Archimedean copula family, and bivariate Gumbel copula density function is
\begin{equation}
	c_{g}(\mathbf{u})=exp\left\{-\left[\sum_{i=1}^{2}(-\ln u_i)^\alpha\right]^{\frac{1}{\alpha}} \right\}\left[\left( \left[\sum_{i=1}^{2}(-\ln u_i)^\alpha\right]^{\frac{1}{\alpha}-1} \right) \left(\sum_{i=1}^2 \frac{(-\ln u_i)^{\alpha}}{u_i}\right) \right],
	\label{eq:gumbelcopula}
\end{equation}
where $\alpha$ is the parameter of Gumbel copula.

So estimating CE of Gumbel copula hypothesis can be done as follows:
\begin{enumerate}
	\item estimate $\alpha$ with the likelihood method \cite{joe2014};
	\item calculate the value of Gumbel copula with \eqref{eq:gumbelcopula};
	\item calculate CE of Gumbel copula with \eqref{eq:celikelihood}.
\end{enumerate}

\section{Simulations}
\label{sec:sim}
We test the proposed method with two simulation experiments \footnote{The code is available at \url{https://github.com/majianthu/tch}}. The first experiment simulate bivariate Gaussian copula with correlation coefficient $\rho$ range from 0.1 to 0.9 by the step 0.1. The second experiment simulate bivariate Gumbel copula with the parameter $alpha$ changing from 2 to 10 under the margins being standard normal distribution and exponential distribution. All the sample size of simulations is 300.

We applied the above testing method of Gaussian copula hypothesis and Gumbel copula hypothesis to simulated data and derived two estimated statistics from each sample set.

\begin{figure}
	\centering
	\includegraphics[width=0.62\textwidth]{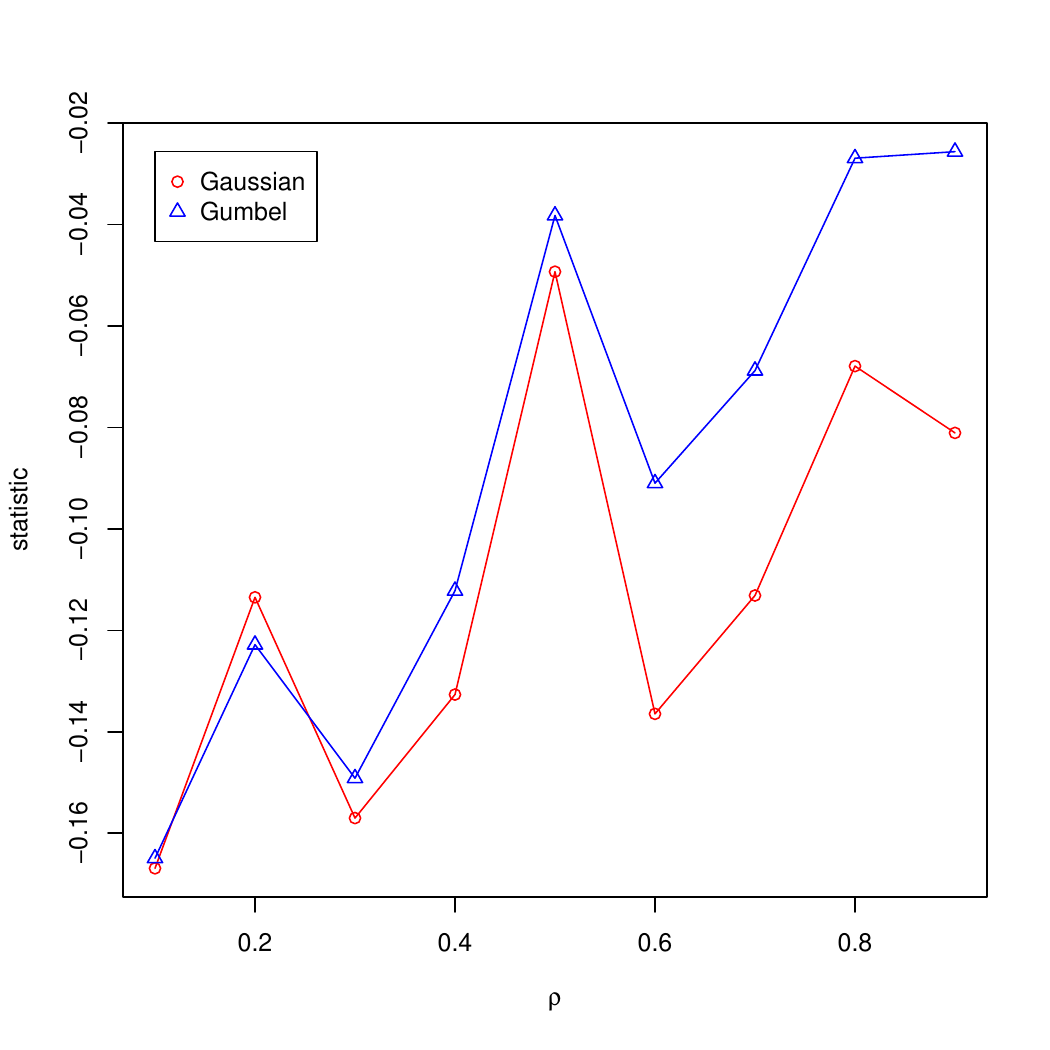}
	\caption{Results of Gaussian copula hypothesis simulation experiments.}
	\label{fig:tgc1}
\end{figure}

\begin{figure}
	\centering
	\includegraphics[width=0.62\textwidth]{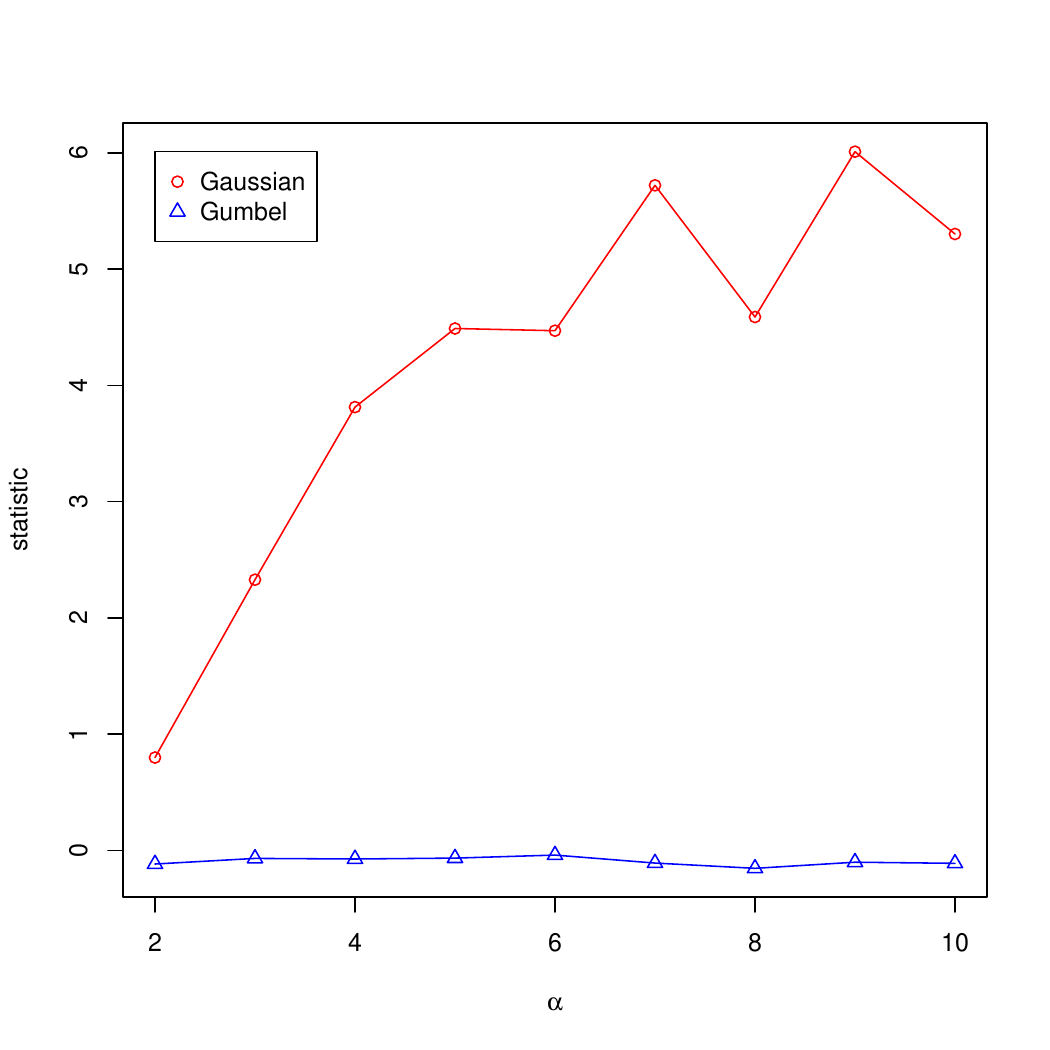}
	\caption{Results of Gumbel copula hypothesis simulation experiments.}
	\label{fig:tgc2}
\end{figure}

The experimental results is shown in Figure \ref{fig:tgc1} and Figure \ref{fig:tgc2}. It can be learned that in the first simulation experiment, the estimated statistics of Gaussian copula hypothesis is smaller than those of Gumbel copula hypothesis and that in the second simulation experiment, the estimated statistics of Gumbel copula hypothesis is smaller than those of Gaussian copula hypothesis. These mean that Gaussian copula hypothesis is true and Gumbel copula hypothesis is true in two experiments respectively.

The \textsf{R} package \texttt{copula} \cite{Yan2007} and \texttt{gumbel} \cite{Dutang2024} were used for Gumbel copula. The \textsf{R} package \texttt{mvtnorm} \cite{Genz2009} waw used for simulating Gaussian distribution. The \textsf{R} package \texttt{copent} \cite{Ma2021} was used as the implementation of CE estimator. 


\section{Conclusions}
\label{sec:con}
In this paper, we proposed a copula hypothesis testing with copula entropy. The test statistic is defined. The estimation method of the proposed statistic is proposed and two special cases of tests for Gaussian copula hypothesis and Gumbel copula hypothesis is given. The effectiveness of the proposed method is verified with simulation experiments.

\bibliographystyle{unsrt}
\bibliography{tch}

\end{document}